# Femtosecond laser direct-write waveplates based on stress-induced birefringence

**BEN MCMILLEN,**[*] **CHRISTOS ATHANASIOU, AND YVES BELLOUARD**

*Galatea Lab, STI/IMT, Ecole Polytechnique Fédérale de Lausanne (EPFL), Rue de la Maladière 71b, CH-2002, Neuchâtel, Switzerland*
*\*ben@mcmillen.eu*

**Abstract:** The use of femtosecond lasers to introduce controlled stress states has recently been demonstrated in silica glass. We use this technique, in combination with chemical etching, to generate and control stress-induced birefringence over a well-defined region of interest, demonstrating direct-write wave plates with precisely tailored retardance levels. This tailoring enables the fabrication of laser-written polarization optics that can be tuned to any wavelength for which silica is transparent and with a clear aperture free of any laser modifications. Using this approach, we achieve sufficient retardance to act as a quarter-wave plate. The stress distribution within the clear aperture is analyzed and modeled, providing a generic template that can be used as a set of design rules for laser-machined polarization devices.

## References and links

1. K. M. Davis, K. Miura, N. Sugimoto, and K. Hirao, "Writing waveguides in glass with a femtosecond laser," Opt. Lett. **21**, 1729–1731 (1996).
2. K. Miura, J. Qiu, H. Inouye, T. Mitsuyu, and K. Hirao, "Photowritten optical waveguides in various glasses with ultrashort pulse laser," Appl. Phys. Lett. **71**, 3329–3331 (1997).
3. Y. Kondo, T. Suzuki, H. Inouye, K. Miura, T. Mitsuyu, and K. Hirao, "Three-dimensional microscopic crystallization in photosensitive glass by femtosecond laser pulses at nonresonant wavelength," Jpn. J. Appl. Phys. **37**, L94 (1998).
4. A. Marcinkevicius, S. Juodkazis, M. Watanabe, M. Miwa, S. Matsuo, H. Misawa, and J. Nishii, "Femtosecond laser-assisted three-dimensional microfabrication in silica," Opt. Lett. **26**, 277–279 (2001).
5. Y. Shimotsuma, P. G. Kazansky, J. Qiu, and K. Hirao, "Self-organized nanogratings in glass irradiated by ultrashort light pulses," Phys. Rev. Lett. **91**, 247405 (2003).
6. K. Minoshima, A. M. Kowalevicz, I. Hartl, E. P. Ippen, and J. G. Fujimoto, "Photonic device fabrication in glass by use of nonlinear materials processing with a femtosecond laser oscillator," Opt. Lett. **26**, 1516–1518 (2001).
7. R. Osellame, S. Taccheo, G. Cerullo, M. Marangoni, D. Polli, R. Ramponi, P. Laporta, and S. De Silvestri, "Optical gain in Er-Yb doped waveguides fabricated by femtosecond laser pulses," Electron. Lett. **38**, 964–965 (2002).
8. H. Zhang, S. M. Eaton, J. Li, and P. R. Herman, "Type II femtosecond laser writing of Bragg grating waveguides in bulk glass," Electron. Lett. **42**, 1223-1224 (2006).
9. T. Meany, M. Gräfe, R. Heilmann, A. Perez-Leija, S. Gross, M. J. Steel, M. J. Withford, and A. Szameit, "Laser written circuits for quantum photonics: Laser written quantum circuits," Laser Photonics Rev. **9**, 363–384 (2015).
10. J. Tang, J. Lin, J. Song, Z. Fang, M. Wang, Y. Liao, L. Qiao, and Y. Cheng, "On-chip tuning of the resonant wavelength in a high-Q microresonator integrated with a microheater," Int. J. Optomechatronics **9**, 187–194 (2015).
11. Y. Bellouard, A. Said, M. Dugan, and P. Bado, "Monolithic three-dimensional integration of micro-fluidic channels and optical waveguides in fused silica," MRS Proceedings, **782**, 63–68 (2003).
12. Y. Cheng, K. Sugioka, and K. Midorikawa, "Microfluidic laser embedded in glass by three-dimensional femtosecond laser microprocessing," Opt. Lett. **29**, 2007–2009 (2004).
13. Y. Hanada, K. Sugioka, I. Shihira-Ishikawa, H. Kawano, A. Miyawaki, and K. Midorikawa, "3D microfluidic chips with integrated functional microelements fabricated by a femtosecond laser for studying the gliding mechanism of cyanobacteria," Lab Chip **11**, 2109–2115 (2011).
14. F. Bragheri, L. Ferrara, N. Bellini, K. C. Vishnubhatla, P. Minzioni, R. Ramponi, R. Osellame, and I. Cristiani, "Optofluidic chip for single cell trapping and stretching fabricated by a femtosecond," J. Biophoton. **3**, 234–243

(2010).
15. A. Schaap, Y. Bellouard, and T. Rohrlack, "Optofluidic lab-on-a-chip for rapid algae population screening," Biomed. Opt. Express **2**, 658–664 (2011).
16. Y. Bellouard, A. Said, and P. Bado, "Integrating optics and micro-mechanics in a single substrate: a step toward monolithic integration in fused silica," Opt. Express **13**, 6635–6644 (2005).
17. L. A. Fernandes, J. R. Grenier, J. S. Aitchison, and P. R. Herman, "Fiber optic stress-independent helical torsion sensor," Opt. Lett. **40**, 657 (2015).
18. B. Lenssen and Y. Bellouard, "Optically transparent glass micro-actuator fabricated by femtosecond laser exposure and chemical etching," Appl. Phys. Lett. **101**, 103503 (2012).
19. T. Yang and Y. Bellouard, "Monolithic transparent 3D dielectrophoretic micro-actuator fabricated by femtosecond laser," J. Micromech. and Microeng. **25**, 105009 (2015).
20. M. Beresna, M. Gecevičius, and P. G. Kazansky, "Polarization sensitive elements fabricated by femtosecond laser nanostructuring of glass [Invited]," Opt. Mater. Express **1**, 783–795 (2011).
21. M. Beresna, M. Gecevicius, P. G. Kazansky, and T. Gertus, "Radially polarized optical vortex converter created by femtosecond laser nanostructuring of glass," Appl. Phys. Lett. **98**, 201101–201101 (2011).
22. K. Mishchik, G. Cheng, G. Huo, I. M. Burakov, C. Mauclair, A. Mermillod-Blondin, A. Rosenfeld, Y. Ouerdane, A. Boukenter, O. Parriaux, and others, "Nanosize structural modifications with polarization functions in ultrafast laser irradiated bulk fused silica," Opt. Express **18**, 24809–24824 (2010).
23. Y. Bellouard, A. Champion, B. Lenssen, M. Matteucci, A. Schaap, M. Beresna, C. Corbari, M. Gecevičius, P. Kazansky, O. Chappuis, M. Kral, R. Clavel, F. Barrot, J. M. Breguet, Y. Mabillard, S. Bottinelli, M. Hopper, C. Hoenninger, E. Mottay, J. Lopez, "The femtoprint project," J. Laser Micro Nanoen. **7**, 1–10 (2012).
24. A. Champion and Y. Bellouard, "Direct volume variation measurements in fused silica specimens exposed to femtosecond laser," Opt. Mater. Express **2**, 789–798 (2012).
25. A. Champion, M. Beresna, P. Kazansky, and Y. Bellouard, "Stress distribution around femtosecond laser affected zones: effect of nanogratings orientation," Opt. Express **21**, 24942–24951 (2013).
26. B. McMillen and Y. Bellouard, "On the anisotropy of stress-distribution induced in glasses and crystals by non-ablative femtosecond laser exposure," Opt. Express **23**, 86–100 (2015).
27. Y. Bellouard, A. Champion, B. McMillen, S. Mukherjee, R. Thomson, Y. Cheng, "Femtosecond laser-induced stress-state inversion and related anisotropies," Optica, under review (2016).
28. C.-E. Athanasiou and Y. Bellouard, "A monolithic micro-tensile tester for investigating silicon dioxide polymorph micromechanics, fabricated and operated using a femtosecond laser," Micromachines **6**, 1365–1386 (2015).
29. Y. Bellouard, "Non-contact sub-nanometer optical repositioning using femtosecond lasers," Opt. Express **23**, 29258 (2015).
30. E. Bricchi, B. G. Klappauf, and P. G. Kazansky, "Form birefringence and negative index change created by femtosecond direct writing in transparent materials," Opt. Lett. **29**, 119–121 (2004).
31. Y. Bellouard, A. Said, M. Dugan, and P. Bado, "Fabrication of high-aspect ratio, micro-fluidic channels and tunnels using femtosecond laser pulses and chemical etching," Opt. Express **12**, 2120–2129 (2004).
32. OpenPolScope [computer software], (2015), http://www.openpolscope.org
33. S. Mehta, M. Shirbak, and R. Oldenbourg, "Polarized light imaging of birefringence and diattenuation at high resolution and high sensitivity," J. Opt. 15, 094007 (2013).
34. S. Rajesh and Y. Bellouard, "Towards fast femtosecond laser micromachining of fused silica: The effect of deposited energy," Optics Express **18**, 21490–21497 (2010).
35. C.-E. Athanasiou and Y. Bellouard, "Investigation of the micro-mechanical properties of femtosecond laser-induced phases in amorphous silica," Proc. SPIE **9740**, 97401E (2016).
36. E. Mathieu, "Théorie de l'Elasticité des Corps Solides," (Gauthier-Villars, Paris, 1890), seconde partie, chap. 10, pp 140 - 178
37. C. Ribière, "Sur divers cas de la flexion des prismes rectangles", thesis Université de Bordeaux, (1889)
38. L. N. G. Filon, "On an approximate solution for the bending of a beam of rectangular cross-section under any system of load, with special reference to points of concentrated or discontinuous loading," Philos. T. Roy. Soc. A, **201**, 63-155 (1903).
39. S. P. Timoshenko, J. N. Goodier, *Theory of Elasticity,* 3$^{rd}$ ed. (McGraw-Hill, 1970), Chap. 3, Sect. 24
40. F. Hecht, "New development in FreeFem++," J. Numer. Math. **20**, 251-266 (2012). http://www.freefem.org
41. A. K. Spilman and T. G. Brown, "Stress birefringent, space-variant wave plates for vortex illumination," Appl. Optics **46**, 61–66 (2007).
42. J. F. Nye, *Physical Properties of Crystals* (Oxford University, 1985).
43. T. N. Vasudevan and R. S. Krishnan, "Dispersion of the stress-optic coefficient in glasses," J. Phys. D Appl. Phys. **5**, 2283 (1972).

## 1. Introduction: femtosecond laser writing as a tool to induce stress in silica

The effect of non-ablative ultrashort pulses on fused silica has been extensively studied over the last decade since the pioneering works that first demonstrated localized increase of refractive index [1,2], etching rate [3,4], and the occurrence of self-organized nanostructures [5]. These laser-induced modifications have been used in numerous applications, including integrated optics [1,6-10], optofluidics [11-15], optomechanics [16,17], actuators [18,19], and polarization optics [20-22], to name a few. A particular feature of this manufacturing process is that it can be implemented with compact, tabletop 3D printing systems [23].

Femtosecond laser exposure in the non-ablative regime induces a confined change of volume [24], creating oriented tensile or compressive stress fields capable of reaching locally high levels of stress (up to ~ 2 GPa in compression) [25]. Interestingly, these laser-induced stress fields can be tailored and controlled by changing the orientation of the laser polarization with respect to the writing direction [25,26], and by tuning the pulse duration of the laser [27]. This not only allows for control of the stress orientation, but also the magnitude and state, i.e. switching from tensile to compressive stress. This ability to fine-tune stress fields finds its origins in the variety of nanostructures induced in laser-modified materials, as highlighted in [25-27]. This intriguing effect has already been applied to tensile testing procedures for investigating glass micromechanics [28], as well as for future concepts of optical component packaging [29].

Here, we explore the use of controlled stress-states for applications in polarization optics. Specifically, we present a technique for manufacturing polarization devices using the indirect action of laser-induced stress. Device geometry is defined using laser-machined silica substrates, limiting the action of the induced stress (and resulting birefringence) to a particular aperture. Unlike other techniques, which rely on direct light interaction with form-birefringent nanogratings [30], this clear aperture is free from laser-modifications, potentially affording higher power handling capabilities, higher overall device transmission, and a cleaner undistorted beam profile. Additionally, the use of fused silica substrates provides a broad transmission window, potentially enabling a wide tuning range for the operating wavelength. With fast fabrication times and full control over the device retardance, broadband-like devices consisting of many waveplates on a single substrate are possible. Furthermore, by using the same femtosecond laser for both machining and subsequent loading, the device manufacturing process can be greatly simplified.

## 2. Waveplate concept

The principle of a laser-machined waveplate is shown in Fig. 1. A rectangular clear aperture (labeled as zone c) is defined by two cuts made through a fused silica substrate using a wet-etch-assisted laser machining process [31].

After etching, a set of lines (which we call 'stressors') are machined in the bulk of the material at opposing ends of the rectangle (b), using writing parameters sufficient to generate nanogratings within the laser-affected zone [5]. These nanogratings are oriented parallel to the laser writing direction, such that the principal component of the stress tensor [25,26] is directed perpendicular to the line orientation.

This arrangement of opposing stressors induces a quasi-uniaxial loading of the material in the center (as indicated by the red arrows in Fig. 1), generating a sizeable optical retardance within the clear aperture. With this simple arrangement, the cuts serve as a stress-free interface, confining the action of the stressors to the area of interest, in which the level of optical retardance can be tailored through careful control of the number, density, and exposure parameters of the stressors. It should be noted that, while the fabrication process presented here is done in multiple steps, the action of machining and device loading could be combined prior to etching, simplifying the production process further.

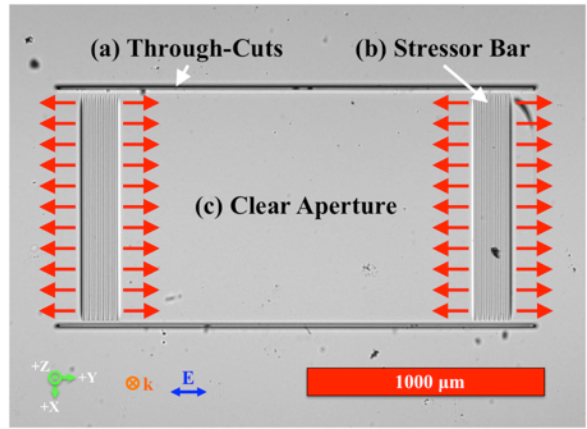

Fig. 1. Microscope image of a laser-machined waveplate. In this arrangement, a clear aperture (c) is defined by a pair of horizontal cuts (a), fabricated using a wet-etch-assisted laser machining process [31]. Stress-induced birefringence is then generated in this aperture by writing a set of vertical lines (labeled 'stressor bar', (b)) within the bulk of the substrate at each end of the device, using exposure conditions sufficient to generate nanogratings. Here, the writing polarization, and hence the orientation of the nanogratings, is chosen such that the generated stress behaves like a quasi-uniaxial loading, as indicated by the red arrows.

## 3.  Experimental procedure

### 3.1 Setup and measurement principle

Waveplates were fabricated in several steps, beginning with the machining of 20 blank devices (no stressors, each device 1x2 mm, staggered on a 4 mm grid) in a 25x25x0.5 mm fused silica substrate, using an amplified femtosecond laser system (Yb-fiber, Amplitude Systèmes), delivering 280 fs pulses at 1030 nm. Here, the laser was focused using a 20x objective (NA 0.4) optimized for 1046 nm. Following machining, the sample was etched in a 2.5% solution of hydrofluoric acid for ~ 8 hours. To maximize etch rate, the variable repetition rate of the laser was set to 760 kHz during machining [31].

Using the same writing configuration, each device was loaded by progressively forming stressors perpendicular to the cuts at both ends of the waveplate. The stressors were then arranged in blocks at each end of the waveplate at a fixed spacing of 5 or 10 μm, with the laser polarization set perpendicular to the writing direction in order to generate maximum stress along the long-axis of the waveplate ($y$-direction in Fig. 1) [25,26]. Here, we define a single 'stressor' as a series of stacked discrete laser-modified regions with a $z$-spacing of 15 μm, forming a modified 'sheet' of material in the $xz$ plane, for a total of 29 lines per sheet.

Each group of stressors was embedded in the bulk of the substrate and spaced 20 μm from all exterior surfaces to limit stress concentration and consequent crack formation. Additionally, each block was inset 100 μm from the ends of the device. To ensure even loading, the repetition rate of the laser was reduced to ~ 80 kHz in order to maintain the required value of energy deposition for maximum stress, while keeping the writing velocity low, thereby limiting the influence of acceleration dynamics of the motion stages used during fabrication.

The resulting stress-induced birefringence was measured using an optical microscope (Olympus BX51) equipped with a liquid-crystal universal compensator for measuring birefringence (VariLC). The LC unit was controlled using freely available software (OpenPolScope [32-33]), which provided all the necessary functions for calibration of the LC compensator and retardance measurement acquisition. Further post-processing in MATLAB was then used to apply a false color-map to each measurement, as well as extract retardance data. For the purposes of this experiment, we only focus on the measured absolute retardance.

## 3.2 Experimental results

We first examine the evolution of stress within the clear aperture of the waveplates as a function of deposited energy [34]. A pulse energy of 250 nJ was selected, corresponding to a laser-modification which lies solidly within the nanograting regime for optimal stress generation [25,26]. The speed was varied from 1700 to 170 μm/s, resulting in a deposited energy range of 10 to 100 J/mm$^2$. The results of this experiment are shown in Fig. 2 below.

For this experiment, the line spacing of individual stressors was fixed at 10 μm, with a total of 16 stressors per side (32 per device). From Fig. 2, the developed stress peaks around 20 J/mm$^2$, giving a maximum retardance of just over 15 nm. Above this point, the retardance begins to decay, which we interpret as a consequence of stress-relaxation due to crack formation within the nanogratings. This measurement not only serves as a calibration to develop maximum stress during the writing process, but also confirms previous findings for stress evolution in the nanograting regime [24]. For reference, the phase shift at 546 nm (the wavelength used by the VariLC measurement system) is given on the right hand side of the graph.

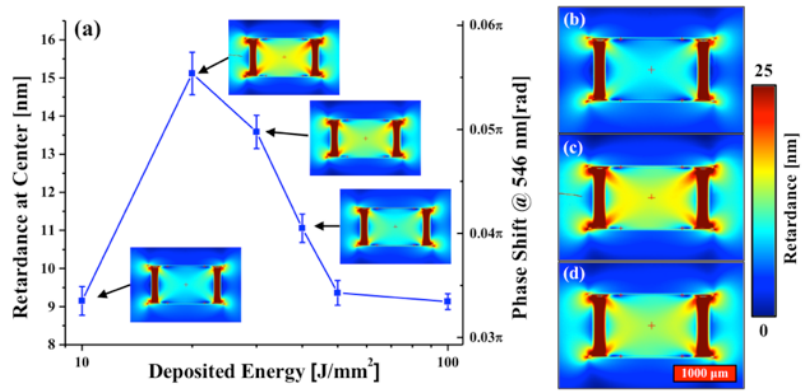

Fig. 2. (a) Retardance as a function of the energy deposited for a pulse energy of 250 nJ and a line spacing of 10 μm. All devices were written with 16 stressors per side (32 stressors total per device). For these writing conditions, it was found that the stress peaks for a deposited energy of 20 J/mm$^2$. The images in (b)-(d) show a magnified view of the retardance map for three waveplates spanning the range over the peak shown in (a), with deposited energy values of 10, 20, and 30 J/mm$^2$ respectively.

While the peak retardance developed in Fig. 2 is only on the order of ~ 15 nm, this plot serves as a calibration for process control to determine the best writing conditions for generating stress in a single line. For further optimization of the developed stress, we now turn our attention to the lateral spacing of the stressors.

Based on the results shown in Fig. 2, a deposited energy target of 20 J/mm$^2$ was chosen to fabricate three additional devices: one with a stressor spacing of 10 μm and 64 total stressors, and two with a reduced spacing of 5 μm, each with 64 and 128 total stressors respectively. The results are shown in Fig. 3 below. The device in Fig. 3(a) is the same displayed in Fig. 2(a), and is used as a reference (20 J/mm$^2$, ~ 15 nm max retardance).

Here we measure the developed retardance at the center of the clear aperture as a function of the spacing and number of lines in each stressor. Note that the retardance scale has a maximum retardance of 100 nm. For the device shown in Fig. 3(d), a maximum retardance of ~ 55 nm is developed. This retardance map will be used as a comparison and calibration for the following section on device modeling.

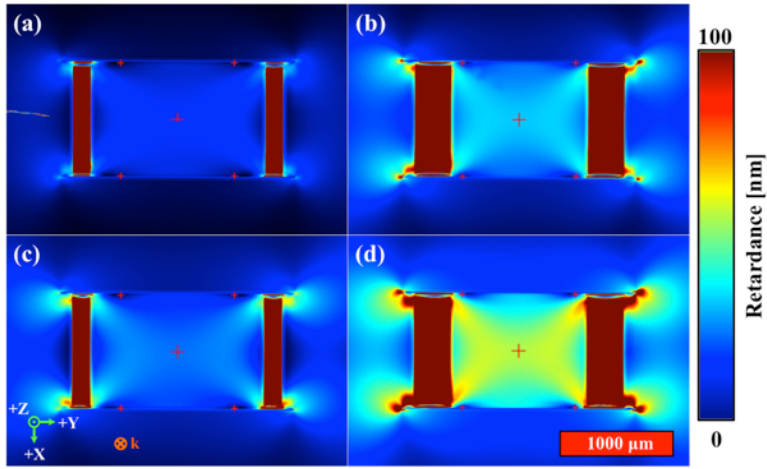

Fig. 3. Waveplate retardance as a function of the number of stressors and stressor spacing. (a) and (b) show retardance maps for a stressor spacing of 10 μm, with 16 and 32 stressors per side, respectively. Devices (c) and (d) are similar, but for a stressor spacing of 5 um and 32 and 64 stressors per side, respectively. All devices were written with a deposited energy of 20 J/mm², with the laser polarization fixed parallel to the long axis of each device. For comparison, the device shown in (a) has a peak center retardance of ~ 15 nm, while the device in (d) is considerably higher at ~ 55 nm.

## 4. Modeling and discussion

We now set out to describe the behavior of the developed stress in the waveplates in terms of a set of design parameters, such as the device geometry and the number and density of the stressors. These models may then be used as a set of design rules, giving full control over the stress field under different geometrical and loading conditions.

*4.1 One dimensional analytical model*

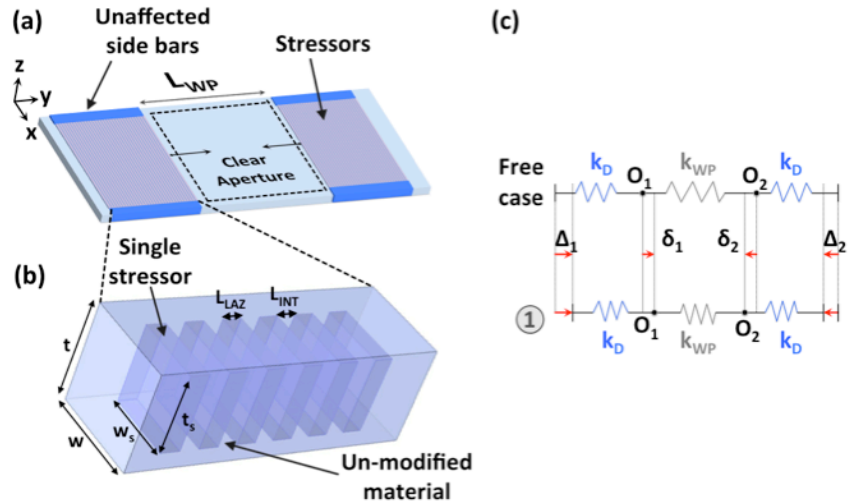

Fig. 4. (a) 3D visualization of a waveplate, highlighting the various regions taken into consideration in the 1D spring model. (b) Expanded view of the stressor region, showing the composite nature of the structure. (c) Lumped spring model used to predict the retardance in the center of the clear aperture as a function of the number of stressors.

As a first approximation, we consider a one-dimensional lumped model (illustrated in Fig. 4 below), following an approach reported in [28]. Here the modified and unmodified regions of the material are represented by springs, for which an equivalent stiffness is derived.

In this model, the geometry of the waveplate is broken down into several distinct regions, as shown in Fig. 4(a). These regions are labeled as the stressors (composite laser modified volume), the un-affected sidebars, and the clear aperture of the waveplate. Each of these regions may then be represented by an equivalent stiffness, as shown in Fig. 4(c).

The model is built by first considering the composite region containing each stressor block, which is treated as a parallel and series combination of springs and assigned an equivalent stiffness, $K_D$, as illustrated in Fig. 4(c). This assembly is then placed in series with the clear aperture of the waveplate, $K_{WP}$.

After formulation of the total stiffness of the system, and taking into account stress-induced birefringence, we arrive at the following expression for calculating the retardance in terms of the number of stressors, $n$:

$$R = C\sigma t \approx \underbrace{(E_{SiO_2})}_{Material} \underbrace{(\varepsilon_{LAZ})}_{Laser} \underbrace{\left[ nt \frac{\alpha}{\left(1 + \frac{L_{WP}}{L_{INT}} \lambda \right)} \right]}_{Geometry} \tag{1}$$

where the factor $\lambda$ is given by:

$$\lambda = \frac{1}{n} \left[ \frac{2(1-\beta)}{\alpha + \left(1 - \frac{1}{n}\right)} + \frac{\beta}{\kappa \alpha \gamma + \left(1 - \frac{1}{n}\right)} \right] \cong \frac{1}{n} \left[ \frac{2(1-\beta)}{\alpha + 1} + \frac{\beta}{\kappa \alpha \gamma + 1} \right] \tag{2}$$

The parameters $\alpha$, $\beta$ and $\gamma$ are dimensionless ratios, defined as $\alpha = L_{LAZ}/L_{INT}$, $\beta = t/t_s$ and $\gamma = w/w_s$, respectively, while $\kappa$ is the ratio of the Young's modulus of pristine SiO$_2$ to that of laser modified SiO$_2$. Note that in previous work [28], we estimated the Young's modulus for the laser-affected zones (nanograting modification regime) to be approximately one third of the modulus of pristine, un-modified silica. The quantities $w_s$, $t_s$, $w$, and $t$ are the width and length of a stressor and surrounding material, respectively, while the quantities $L_{INT}$ and $L_{LAZ}$ correspond to the spacing between laser-modified regions (vertical sheets) and the thickness of these regions (along the $y$-direction), respectively. $C$ is the stress-optic coefficient (see Appendix), $L_{WP}$ is the initial length of the clear-aperture, $E_{SiO2}$ is the elastic modulus of fused silica, and $\varepsilon_{LAZ}$ is the net volumetric expansion induced during laser exposure [25].

From a design point of view, Eq. (1) gives an overview of the tuning parameters available to the user. The choice of Young's modulus is fixed by the choice of fused silica (material component), while the induced strain (laser-induced component) is also relatively fixed for optimized stress generation, as demonstrated by the waveplates shown in Fig. 3. This leaves device geometry as the main parameter, and we note that, for large numbers of stressors, the parameter $\lambda$ converges to zero, leaving only the ratio of the stressor spacing, material thickness, and the number of stressors.

It should be noted that this model is a simplification, as the interfacial energy between the various domains (laser-affected, pristine, etc.) is not considered. However, so long as we assume that the behavior of these regions is governed by linear elasticity, this approximation remains valid. Despite these subtleties, the mechanical behavior of the system is reasonably well predicted using this approach.

In addition to the simplifications mentioned, we make the following assumptions:

1/ Fine structures within the laser-modified composite are ignored, and instead treated as a homogeneous material with a lower density than the surrounding, unmodified bulk. An average elastic modulus of 34 GPa was chosen for this region, which was recently reported in [28].

2/ For the particular arrangement of embedded stressors used here, a certain portion of the device loading is transferred to the bulk as shear stress; in this case along the regions label 'unaffected sidebars'. We consider this shear low and negligible, as calculated based the shear-lag model used in [28].

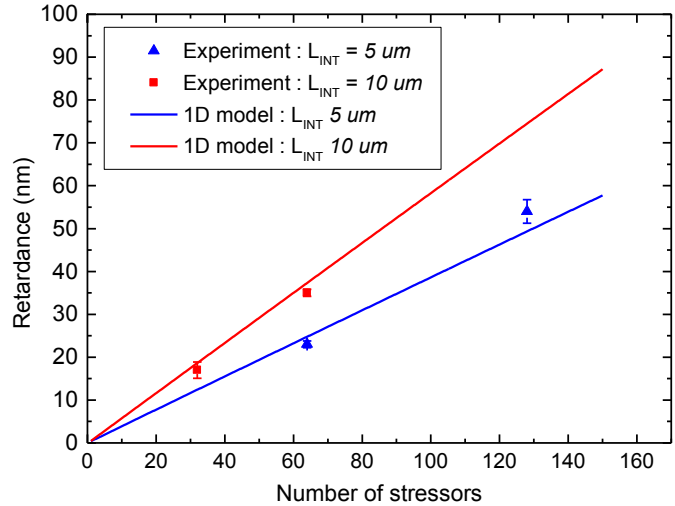

Fig. 5. Prediction of waveplate retardance as a function of the number and spacing of machined stressors using the 1D stiffness model outlined in section 4.1. Here, the individual data points the developed retardance in each waveplate (measured in the center of the clear aperture for each device shown in Fig. 3), while the curves are the predicted values based on the model. For each measured data point, the error was estimated from the measurement noise, which on average was not greater than ± 1 nm. The reader should note that the measured data points shown in the above graph compare different sizes of clear aperture. The devices written with 32 stressors 10 μm spacing and 64 stressors 5 μm spacing have the same clear aperture size, while conversely the second two data points are also comparable.

Figure 5 plots the developed retardance as a function of the number of stressors, comparing the predicted retardance based on the 1D model with the measured values from the center of each waveplate shown in Fig. 3.

Overall, the model is in good agreement for low and high numbers of stressors, however there is some deviation for higher numbers of stressors. While this model provides a reasonable estimation, the accuracy it provides is limited, as we do not consider the 2D interaction of stress present in the waveplate. For an improved model, we now examine the full 2D structure, which we cover in the next section.

*4.2 Two-dimensional analytical model*

While the simplified 1D model discussed in the previous section provides an adequate approach to predicting stress evolution from the number and density of machined stressors, it does little to describe the full contour of the induced retardance field. For analytical derivation of this full contour, we use a more refined model based on Airy stress functions [36-39]. In this model, the waveplate is treated as a two-dimensional body (finite rectangular domain) with a continuously distributed load along the width (provided by the action of the stressors), as depicted in Fig. 6.

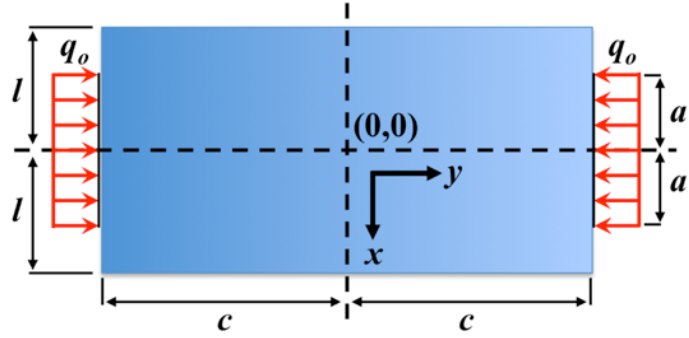

Fig. 6. Schematic representation of the loading case considered for the two-dimensional model. The terms here represent the dimensions of the waveplate ($c$, $l$), as well as the applied load ($q_o$) and the area over which it is distributed ($a$).

While a full derivation on the use of Airy stress functions is beyond the scope of this work, we will briefly outline the process here, referring to the notation used in Fig. 6. Here, we assume a right-handed coordinate system, the origin of which is taken to be in the center of the rectangle defining the clear aperture. Following the work in [39], a general solution to the stress developed in the waveplate is found by defining a stress function $\varphi(x,y)$, which satisfies the biharmonic equation:

$$\frac{\partial^4 \varphi}{\partial x^4} + 2\frac{\partial^4 \varphi}{\partial x^2 \partial y^2} + \frac{\partial^4 \varphi}{\partial y^4} = 0. \tag{3}$$

The equation $\varphi(x,y)$ is of the form:

$$\varphi(x,y) = \sin\frac{m\pi x}{l} f(y) \tag{4}$$

where $m$ is an integer, and $y$ is the only independent variable of $f(y)$. Substituting Eq. (4) into Eq. (3), and making the substitution $\alpha = m\pi/l$, leads to the following equation for determining $f(y)$:

$$\sin(\alpha x)\left[f^{(4)}(y) - 2\alpha^2 f^{(2)}(y) + \alpha^4 f(y)\right] = 0. \tag{5}$$

The stress function then becomes, after integration of Eq. (5):

$$\varphi(x,y) = \sin(\alpha x)\left[C_1 \cosh(\alpha y) + C_2 \sinh(\alpha y) + C_3 y \cosh(\alpha y) + C_4 y \sinh(\alpha y)\right] \tag{6}$$

where the coefficients $C_1 - C_4$ are computed using the boundary conditions of the substrate at the upper and lower edges ($y = \pm c$). In order to accommodate generalized representations of the load applied to one edge of the substrate, a Fourier series is used, as follows:

$$q(x) = A_o + \sum_{m=1}^{\infty} A_m \sin(\alpha x) + \sum_{m=1}^{\infty} A'_m \cos(\alpha x), \tag{7}$$

with an additional loading term computed for the lower edge. For our case of symmetrical loading, the terms containing $sin(\alpha x)$ vanish from Eq. (7), and the coefficients $A_o$ and $A'_m$ may be computed by:

$$A_o = \frac{q_o a}{l}, \quad A'_m = \frac{1}{l}\int_{-a}^{a} q_o \cos(\alpha x) \partial x = \frac{2q_o}{m\pi}\sin(\alpha a). \tag{8}$$

Here, $A_o$ represents the uniform load applied to the right-hand edge. Finally, the 2D stress tensor is related to the stress function by:

$$\sigma_{xx} = \frac{\partial^2 \varphi}{\partial y^2}, \; \sigma_{yy} = \frac{\partial^2 \varphi}{\partial x^2}, \; and \; \tau_{xy} = \frac{\partial^2 \varphi}{\partial x \partial y} \tag{9}$$

where $\sigma_{xx}$, $\sigma_{yy}$, and $\tau_{xy}$ are the *x*, *y*, and shear components of stress, respectively. Computing the derivatives of Eq. (6), we arrive at the solutions that satisfy the biharmonic equation, giving a full tensorial description of the stress field in the clear aperture of the waveplate:

$$\sigma_{xx} = \frac{q_o a}{l} + \frac{4q_o D}{\pi} \sum_{m=1}^{\infty} \frac{\sin(\alpha a)}{m} \frac{[\alpha c \, \text{ch}(\alpha c) - \text{sh}(\alpha c)] \text{ch}(\alpha y) - \alpha y \, \text{sh}(\alpha y) \text{ch}(\alpha c)}{\text{sh}(2\alpha c) + 2\alpha c} \cos(\alpha x)$$

$$\sigma_{yy} = -\frac{q_o a}{l} - \frac{4q_o D}{\pi} \sum_{m=1}^{\infty} \frac{\sin(\alpha a)}{m} \frac{[\alpha c \, \text{ch}(\alpha c) + \text{sh}(\alpha c)] \text{ch}(\alpha y) - \alpha y \, \text{sh}(\alpha y) \text{ch}(\alpha c)}{\text{sh}(2\alpha c) + 2\alpha c} \cos(\alpha x) \tag{10}$$

$$\tau_{xy} = \frac{q_o a}{l} + \frac{4q_o D}{\pi} \sum_{m=1}^{\infty} \frac{\sin(\alpha a)}{m} \frac{\alpha c \, \text{ch}(\alpha c) \text{sh}(\alpha y) - \alpha y \, \text{ch}(\alpha y) \text{sh}(\alpha c)}{\text{sh}(2\alpha c) + 2\alpha c} \sin(\alpha x)$$

with $\alpha = \frac{m\pi}{l}$, $ch = \cosh$, and $sh = \sinh$.

Finally, using the stress components shown in Eq. (10), the birefringence field was computed using the photoelastic relationships outlined in the Appendix. Note that, in our calculations we ignore the contribution from shear stress, as this component plays a minimal role in the final device retardance. Finally, for the equations shown above, we compute the stress components for the first fifty orders of the Fourier series, which we find adequately describes the stress-field developed in the clear aperture. The coefficient *D* is a correction factor, which will be discussed in Section 4.4.

### *4.3 Finite element modeling*

Finite element modeling was used as a means to verify the experimental results, as well as to provide an additional metric for comparison against the 2D analytical model discussed in the previous section. Of the many FEM software packages available, freeFEM++ [40] was chosen due to its low-level programmatic approach, giving a high degree of control over the modeling process.

The simulation domain was modeled after the peak device shown in Fig. 3(d) (maximum center retardance of ~ 55 nm) using the displacement formulation [40]. As with the 1 and 2D analytical models, the stressors were treated as a homogeneous volume, and in this model, were represented as a rectangular region occupying the same area as the stressors of the original device. Furthermore, Dirichlet boundary conditions were used, specifying that all displacements must go to zero at the outer boundaries of the substrate. For this reason, the outer boundaries were grounded while all other boundaries internal to the domain were free to move.

In order to simulate the loading action of the stressors, an initial displacement was specified, directed along the long dimension of the device (see Fig. 1), for both lateral boundaries of each rectangle representing the stressor region. This value was computed by multiplying the size of the stressor region along the *y*-direction (here approximately 315 μm) by a factor of 0.03%, the average unit strain for nanograting-induced stress (experimentally measured in [24]).

Finally, stress-induced birefringence was calculated based on the photoelastic equation (outlined in the Annex), and the resulting field distribution was exported for processing in MATLAB. The results of this simulation, showing only the region of the clear aperture, are shown in Fig. 7.

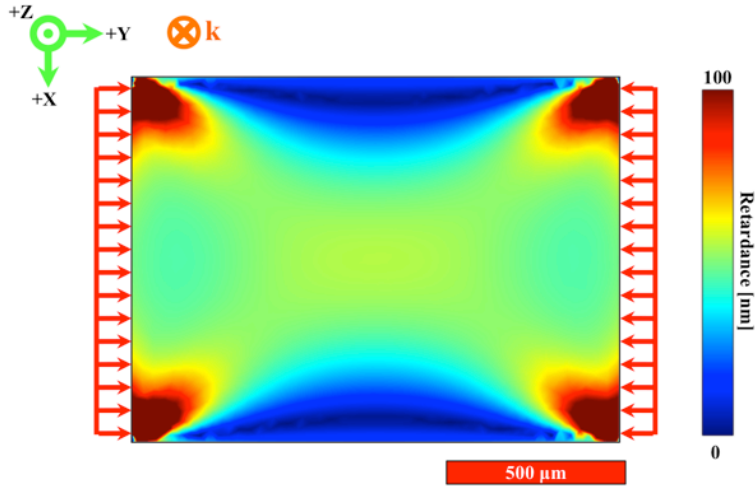

Fig. 7. Simulated retardance profile of a laser-written waveplate using a computed displacement on the boundaries of the stressors. The resulting average strain required to produce a maximum center retardance of 55 nm was ~ 0.16 %. For clarity, the loading along the edges of the plate is shown by red arrows.

It should be noted that the value 0.03 % is an *average* strain for nanograting-induced stress, and as we have done here, was computed by treating the composite region of the laser-modified material as homogeneous. As such, the resulting calculated retardance in the FEM model (when using this value) is under-estimated. To achieve a retardance value closely matching that of the original device, an iterative approach was used, gradually increasing the 'virtual' strain in the model until the desired retardance at the center of the clear aperture was achieved, resulting in a final value of ~ 0.16 %.

### 4.3 Comparison with experimental results

Figure 8 illustrates a comparison between experimental data, FEM simulation, and the analytical model for line profiles taken along the *x* and *y* directions through the center of the clear aperture. For reference, the retardance map for the peak device in Fig. 3(d) is shown in (a).

We first examine the prediction of the shape of the stress field, as illustrated by the contours in Fig. 8(e-g). Qualitatively, the 2D model (f) is in good agreement with the contour of the measured retardance (e), while the FEM model (g) has similar features, but deviates near the boundaries. While the 2D model performs well for the contour, the overall magnitude variation is not properly estimated, as shown by the line profile plots for the *y* and *x* directions in (c) and (d) respectively. The upper insets show the analytical model prediction, and while the shape of the profile closely follows the measurement, the overall magnitude is incorrect.

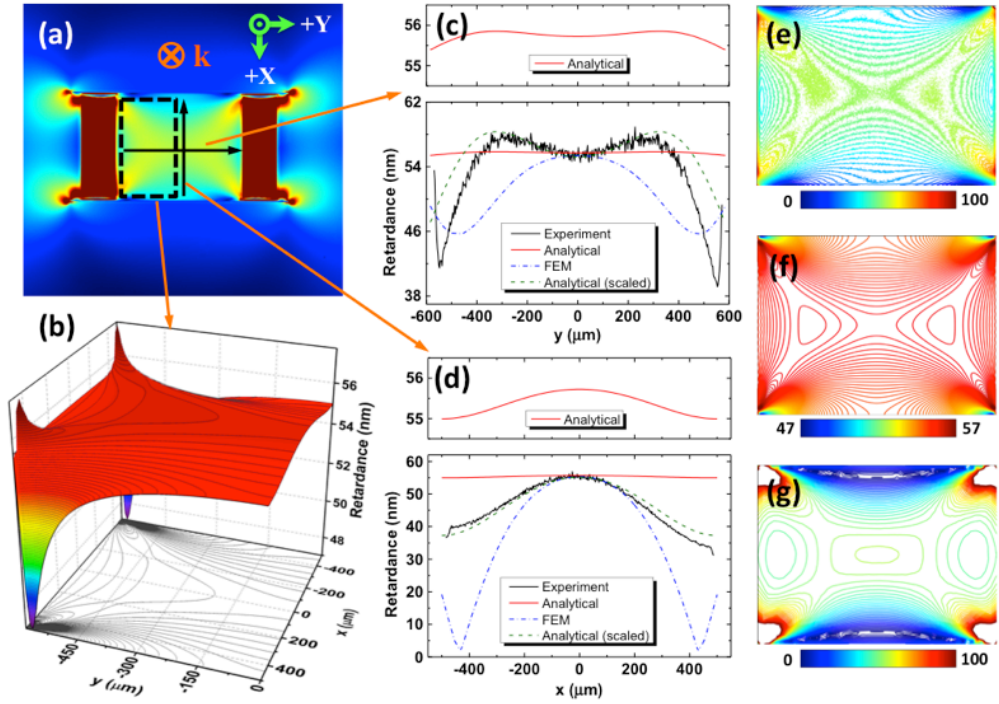

Fig. 8. (a) Retardance map for the peak device shown in Fig. 3(d) (55 nm, 128 stressors, 5 μm spacing). (b) 3D view of the retardance profile calculated using the 2D analytical model (only half the computed map is shown). (c-d) comparison between retardance data obtained for experimental measurements, FEM simulation, and the 2D analytical model for the $y$ and $x$ profile directions respectively. (e-g) Contour maps of the measured retardance, 2D analytical model, and FEM simulation respectively.

We suspect that this may be caused by a slight out-of-plane deformation, suggesting that the loading induced by the stressors requires further optimization. Indeed, no compensation for spherical aberration was used during the writing process. This can account for a decrease of localized intensity with increased writing depth, resulting in a slightly unequal distribution of stress across the substrate thickness. This stress gradient generates a net moment, and in turn favors out-of-plane bending. Such effects have been reported in previous work [28] and can be easily corrected, either by adjusting the power or the stressor spacing with respect to depth within the sample. To adjust the analytical model, we have included a scaling factor (labeled as the variable $D$ in Eqs. (10)), which is shown here for a value of 25, indicating that the overall prediction of the 2D model is in agreement with the measurement, provided that this adjustment is made. We call the reader's attention to the fact that this is an artifact for curve fitting. The exact physical meaning of the parameter $D$ in terms of stress remains unclear at this stage.

The FEM model also deviates considerably from the measurement, which may be attributed to non-uniformity of the distributed load on the as-measured device over the thickness close to the boundaries of the applied load.

## 5. Conclusion

This work demonstrates the use of direct-write femtosecond laser processing to induce predictable, controlled stress states within a chosen region of a substrate through the indirect action of stress-induced birefringence. Here, we utilize this technique to fabricate a polarization device through the indirect action of stress-induced birefringence, resulting in a simple waveplate. In this example, a basic geometry is used, consisting of a series of vertical, laser-written planes arranged in a perpendicular fashion, creating a rectangular region under uniaxial stress.

Though the results presented here are not fully optimized, we have provided a generalized framework, with which further development can be made. The flexibility of the direct-write process allows for virtually infinite combinations of cuts and stressors, opening up new design opportunities for more complex polarization devices with clear apertures and potentially zero transmission losses. For example, one could consider devices based on two-dimensional stress states (as demonstrated in [41] at the macro-scale), leading to the creation of optical vortices. The technique we have introduced here allows for such a concept to be implemented at the micro-scale, and may be tailored to a variety of complex stress states.

**Annex: Photelasticity and principal stress**

The measured retardance is a result of a change in dielectric permittivity due to applied stress, and is given by [42]:

$$\zeta_{ij} = \pi_{ijkl}\sigma_{kl} \quad \text{where} \quad i,j,k,l = 1,2,3 \tag{11}$$

where $\pi_{ijkl}$ is the fourth-rank piezo-optic tensor. Using Voigt notation, the tensorial equation in matrix form for a biaxial planar stress in an isotropic material becomes:

$$\zeta_{(1..6)} = \begin{bmatrix} \pi_{11} & \pi_{12} & \pi_{12} & 0 & 0 & 0 \\ \pi_{12} & \pi_{11} & \pi_{12} & 0 & 0 & 0 \\ \pi_{12} & \pi_{12} & \pi_{11} & 0 & 0 & 0 \\ 0 & 0 & 0 & \pi_{14} & 0 & 0 \\ 0 & 0 & 0 & 0 & \pi_{14} & 0 \\ 0 & 0 & 0 & 0 & 0 & \pi_{14} \end{bmatrix} \begin{bmatrix} \sigma_1 \\ \sigma_2 \\ 0 \\ 0 \\ 0 \\ 0 \end{bmatrix} = \begin{bmatrix} \pi_{11}\sigma_1 + \pi_{12}\sigma_2 \\ \pi_{12}\sigma_1 + \pi_{11}\sigma_2 \\ \pi_{12}\sigma_1 + \pi_{12}\sigma_2 \\ 0 \\ 0 \\ 0 \end{bmatrix} \tag{12}$$

where the birefringence ellipsoid ($\Delta n$) view from the third axis (here the optical axis) is given in terms of $\zeta$ as:

$$(\Delta n)_3 = \left|\left(n_o - \frac{n_o^3}{2}\zeta_1\right) - \left(n_o - \frac{n_o^3}{2}\zeta_2\right)\right| \tag{13}$$

or,

$$(\Delta n)_3 = \frac{n_o^3}{2}\left|(\sigma_1 - \sigma_2)(\pi_{11} - \pi_{12})\right| \tag{14}$$

and finally the retardance is given by:

$$R = C|\sigma_1 - \sigma_2|t \quad \text{and} \quad C = \frac{n_o^3}{2}(\pi_{11} - \pi_{22}) \tag{15}$$

where $t$ is the thickness of the substrate, the value for $C$, the stress-optic coefficient for fused silica, is 3.55 $10^{-12}$ Pa$^{-1}$ [43].

Note that the above equations rely on the principal stresses found in the substrate, which for the case of the FEM and 2D analytical models, must be computed from the $x$, $y$, and shear stress components using the following relation:

$$\sigma_1, \sigma_2 = \frac{\sigma_{xx} + \sigma_{yy}}{2} \pm \sqrt{\left(\frac{\sigma_{xx} - \sigma_{yy}}{2}\right)^2 + \tau_{xy}^2} \tag{16}$$

**Acknowledgements**

BMcM performed the experiments reported in this paper and the FEM modeling. CA derived the analytical models. YB proposed the concept of waveplates presented in this paper and supervised the research. All authors contributed to the writing and proof-reading.

**Funding**

This work is supported by the European Research Council (ERC-2012-StG-307442, 'Galatea'). The Galatea lab acknowledges the sponsoring of Richemont International.